\documentstyle[12pt]{article}
\newcommand{\nc}{\newcommand}
\nc{\renc}{\renewcommand}

\nc{\etal}{\mbox{\it et al. }}
\nc{\ie}{{\it i.e.}}
\nc{\eg}{{\it e.g.}}

\renc{\thefootnote}{\arabic{footnote}}
\nc{\capt}[1]{{\bf Figure.} {\small\sl #1}}

\nc{\eqs}[2]{\mbox{Eqs.~(\ref{#1},\,\ref{#2})}}
\nc{\eq}[1]{\mbox{Eq.~(\ref{#1})}}

\nc{\figs}[2]{\mbox{Figs.~(\ref{#1},\,\ref{#2})}}
\nc{\fig}[1]{\mbox{Fig~.(\ref{#1})}}

\nc{\tag}[1]{\label{#1} \marginpar{{\footnotesize #1}}}
\nc{\mtag}[1]{\label{#1} \mbox{\marginpar{{\footnotesize #1}}}}
\renc{\baselinestretch}{1.2}
\newlength{\overeqskip}
\newlength{\undereqskip}
\nc{\be}[1]{\begin{equation} \mbox{$\label{#1}$}}
\nc{\bea}[1]{\begin{eqnarray} \mbox{$\label{#1}$}}
\nc{\Section}[2]{\section{#2}\label{#1}}
\nc{\Bibitem}[1]{\bibitem{#1}}
\nc{\Label}[1]{\label{#1}}

\nc{\eea}{\vspace{\undereqskip}\end{eqnarray}}
\nc{\ee}{\vspace{\undereqskip}\end{equation}}
\nc{\bdm}{\begin{displaymath}}
\nc{\edm}{\end{displaymath}}
\nc{\dpsty}{\displaystyle}
\nc{\bc}{\begin{center}}
\nc{\ec}{\end{center}}
\nc{\ba}{\begin{array}}
\nc{\ea}{\end{array}}
\nc{\bab}{\begin{abstract}}
\nc{\eab}{\end{abstract}}
\nc{\btab}{\begin{tabular}}
\nc{\etab}{\end{tabular}}
\nc{\bit}{\begin{itemize}}
\nc{\eit}{\end{itemize}}
\nc{\ben}{\begin{enumerate}}
\nc{\een}{\end{enumerate}}
\nc{\bfig}{\begin{figure}}
\nc{\efig}{\end{figure}}
\nc{\arreq}{&\!=\!&}
\nc{\arrmi}{&\!-\!&}
\nc{\arrpl}{&\!+\!&}
\nc{\arrap}{&\!\!\!\approx\!\!\!&}
\nc{\non}{\nonumber\\*}
\nc{\align}{\!\!\!\!\!\!\!\!&&}

\def\lsim{\; \raise0.3ex\hbox{$<$\kern-0.75em
      \raise-1.1ex\hbox{$\sim$}}\; }
\def\gsim{\; \raise0.3ex\hbox{$>$\kern-0.75em
      \raise-1.1ex\hbox{$\sim$}}\; }
\nc{\DOT}{\hspace{-0.08in}{\bf .}\hspace{0.1in}}
\nc{\Laada}{\hbox {$\sqcap$ \kern -1em $\sqcup$}}
\nc\loota{{\scriptstyle\sqcap\kern-0.55em\hbox{$\scriptstyle\sqcup$}}}
\nc\Loota{{\sqcap\kern-0.65em\hbox{$\sqcup$}}}
\nc\laada{\Loota}
\nc{\qed}{\hskip 3em \hbox{\BOX} \vskip 2ex}

\nc{\real}{{\rm I \! R}}
\nc{\Z}{{\sf Z \!\!\! Z}}
\nc{\complex}{{\rm C\!\!\! {\sf I}\,\,}}
\def\bigid{\leavevmode\hbox{\small1\kern-3.8pt\normalsize1}}
\def\id{\leavevmode\hbox{\small1\kern-3.3pt\normalsize1}}
\nc{\slask}{\!\!\!/}
\nc{\bis}{{\prime\prime}}
\nc{\pa}{\partial}
\nc{\na}{\nabla}
\nc{\ra}{\rangle}
\nc{\la}{\langle}
\nc{\goto}{\rightarrow}
\nc{\swap}{\leftrightarrow}

\nc{\EE}[1]{ \mbox{$\cdot10^{#1}$} }
\nc{\abs}[1]{\left|#1\right|}
\nc{\at}[2]{\left.#1\right|_{#2}}
\nc{\norm}[1]{\|#1\|}
\nc{\abscut}[2]{\Abs{#1}_{\scriptscriptstyle#2}}
\nc{\vek}[1]{{\rm #1}}
\nc{\integral}[2]{\int\limits_{#1}^{#2}}
\nc{\inv}[1]{\frac{1}{#1}}
\nc{\dd}[2]{{{\partial #1}\over{\partial #2}}}
\nc{\ddd}[2]{{{{\partial}^2 #1}\over{\partial {#2}^2}}}
\nc{\dddd}[3]{{{{\partial}^2 #1}\over
	{\partial #2 \partial #3}}}
\nc{\dder}[2]{{{d #1}\over{d #2}}}
\nc{\ddder}[2]{{{d^2 #1}\over{d {#2}^2}}}
\nc{\dddder}[3]{{d^2 #1}\over
	{d #2 d #3}}
\nc{\dx}[1]{d\,^{#1}x}
\nc{\dy}[1]{d\,^{#1}y}
\nc{\dz}[1]{d\,^{#1}z}
\nc{\dl}[1]{\frac{d\,^{#1}l}{(2\pi)^{#1}}}
\nc{\dk}[1]{\frac{d\,^{#1}k}{(2\pi)^{#1}}}
\nc{\dq}[1]{\frac{d\,^{#1}q}{(2\pi)^{#1}}}

\nc{\cc}{\mbox{$c.c.$ }}
\nc{\hc}{\mbox{$h.c.$ }}
\nc{\cf}{cf.\ }
\nc{\erfc}{{\rm erfc}}
\nc{\Tr}{{\rm Tr\,}}
\nc{\tr}{{\rm tr\,}}
\nc{\pol}{{\rm pol}}
\nc{\sign}{{\rm sign}}
\nc{\bfT}{{\bf T }}

\nc{\cA}{{\cal A}}
\nc{\cB}{{\cal B}}
\nc{\cD}{{\cal D}}
\nc{\cE}{{\cal E}}
\nc{\cG}{{\cal G}}
\nc{\cH}{{\cal H}}
\nc{\cL}{{\cal L}}
\nc{\cO}{{\cal O}}
\nc{\cT}{{\cal T}}
\nc{\cN}{{\cal N}}
\nc{\rvac}[1]{|{\cal O}#1\rangle}
\nc{\lvac}[1]{\langle{\cal O}#1|}
\nc{\rvacb}[1]{|{\cal O}_\beta #1\rangle}
\nc{\lvacb}[1]{\langle{\cal O}_\beta #1 |}
\nc{\bb}{\bar{\beta}}
\nc{\bt}{\tilde{\beta}}
\nc{\ctH}{\tilde{\cal H}}
\nc{\chH}{\hat{\cal H}}
\nc{\1}{\aa}
\nc{\2}{\"{a}}
\nc{\3}{\"{o}}
\nc{\4}{\AA}
\nc{\5}{\"{A}}
\nc{\6}{\"{O}}
\nc{\al}{\alpha}
\nc{\g}{\gamma}
\nc{\Del}{\Delta}
\nc{\e}{\epsilon}
\nc{\eps}{\epsilon}
\nc{\lam}{\lambda}
\nc{\om}{\omega}
\nc{\Om}{\Omega}
\nc{\ve}{\varepsilon}
\nc{\mn}{{\mu\nu}}
\nc{\k}{\kappa}
\nc{\vp}{\varphi}

\nc{\advp}[3]{{\it  Adv.\ in\ Phys.\ }{{\bf #1} {(#2)} {#3}}}
\nc{\annp}[3]{{\it  Ann.\ Phys.\ (N.Y.)\ }{{\bf #1} {(#2)} {#3}}}
\nc{\apl}[3]{{\it  Appl. Phys. Lett. }{{\bf #1} {(#2)} {#3}}}
\nc{\apj}[3]{{\it  Ap.\ J.\ }{{\bf #1} {(#2)} {#3}}}
\nc{\apjl}[3]{{\it  Ap.\ J.\ Lett.\ }{{\bf #1} {(#2)} {#3}}}
\nc{\app}[3]{{\it Astropart.\ Phys.\ }{{\bf #1} {(#2)} {#3}}}  
\nc{\cmp}[3]{{\it  Comm.\ Math.\ Phys.\ }{{ \bf #1} {(#2)} {#3}}}
\nc{\cqg}[3]{{\it  Class.\ Quant.\ Grav.\ }{{\bf #1} {(#2)} {#3}}}
\nc{\epl}[3]{{\it  Europhys.\ Lett.\ }{{\bf #1} {(#2)} {#3}}}
\nc{\ijmp}[3]{{\it Int.\ J.\ Mod.\ Phys.\ }{{\bf #1} {(#2)} {#3}}}
\nc{\ijtp}[3]{{\it Int.\ J.\ Theor.\ Phys.\ }{{\bf #1} {(#2)} {#3}}}
\nc{\jmp}[3]{{\it  J.\ Math.\ Phys.\ }{{ \bf #1} {(#2)} {#3}}}
\nc{\jpa}[3]{{\it  J.\ Phys.\ A\ }{{\bf #1} {(#2)} {#3}}}
\nc{\jpc}[3]{{\it  J.\ Phys.\ C\ }{{\bf #1} {(#2)} {#3}}}
\nc{\jap}[3]{{\it J.\ Appl.\ Phys.\ }{{\bf #1} {(#2)} {#3}}}
\nc{\jpsj}[3]{{\it J.\ Phys.\ Soc.\ Japan\ }{{\bf #1} {(#2)} {#3}}}
\nc{\lmp}[3]{{\it Lett.\ Math.\ Phys.\ }{{\bf #1} {(#2)} {#3}}}
\nc{\mpl}[3]{{\it  Mod.\ Phys.\ Lett.\ }{{\bf #1} {(#2)} {#3}}}
\nc{\ncim}[3]{{\it  Nuov.\ Cim.\ }{{\bf #1} {(#2)} {#3}}}
\nc{\np}[3]{{\it  Nucl.\ Phys.\ }{{\bf #1} {(#2)} {#3}}}
\nc{\pr}[3]{{\it Phys.\ Rev.\ }{{\bf #1} {(#2)} {#3}}}
\nc{\pra}[3]{{\it  Phys.\ Rev.\ A\ }{{\bf #1} {(#2)} {#3}}}
\nc{\prb}[3]{{\it  Phys.\ Rev.\ B\ }{{{\bf #1} {(#2)} {#3}}}}
\nc{\prc}[3]{{\it  Phys.\ Rev.\ C\ }{{\bf #1} {(#2)} {#3}}}
\nc{\prd}[3]{{\it  Phys.\ Rev.\ D\ }{{\bf #1} {(#2)} {#3}}}
\nc{\prl}[3]{{\it Phys\ Rev.\ Lett.\ }{{\bf #1} {(#2)} {#3}}}
\nc{\pl}[3]{{\it  Phys.\ Lett.\ }{{\bf #1} {(#2)} {#3}}}
\nc{\prep}[3]{{\it Phys\. Rep.\ }{{\bf #1} {(#2)} {#3}}}
\nc{\prsl}[3]{{\it Proc.\ R.\ Soc.\ London\ }{{\bf #1} {(#2)} {#3}}}
\nc{\ptp}[3]{{\it  Prog.\ Theor.\ Phys.\ }{{\bf #1} {(#2)} {#3}}}
\nc{\ptps}[3]{{\it  Prog\ Theor.\ Phys.\ suppl.\ }{{\bf #1} {(#2)} {#3}}}
\nc{\physa}[3]{{\it  Physica\ A\ }{{\bf #1} {(#2)} {#3}}}
\nc{\physb}[3]{{\it  Physica\ B\ }{{\bf #1} {(#2)} {#3}}}
\nc{\phys}[3]{{\it Physica\ }{{\bf #1} {(#2)} {#3}}}
\nc{\rmp}[3]{{\it  Rev.\ Mod.\ Phys.\ }{{\bf #1} {(#2)} {#3}}}
\nc{\rpp}[3]{{\it Rep.\ Prog.\ Phys.\ }{{\bf #1} {(#2)} {#3}}}
\nc{\sjnp}[3]{{\it Sov.\ J.\ Nucl.\ Phys.\ }{{\bf #1} {(#2)} {#3}}}
\nc{\spjetp}[3]{{\it Sov.\ Phys.\ JETP\ }{{\bf #1} {(#2)} {#3}}}
\nc{\yf}[3]{{\it Yad.\ Fiz.\ }{{\bf #1} {(#2)} {#3}}}
\nc{\zetp}[3]{{\it Zh.\ Eksp.\ Teor.\ Fiz.\  }{{\bf #1}  {(#2)} {#3}}}
\nc{\zp}[3]{{\it Z.\ Phys.\ }{{\bf #1} {(#2)} {#3}}}
\nc{\ibid}[3]{{\sl ibid.\ }{{\bf #1} {#2} {#3}}}
\nc{\rf}[1]{(\ref{#1})}
\nc{\nn}{\nonumber \\*}
\nc{\bfB}{\bf{B}}
\nc{\bfv}{\bf{v}}
\nc{\bfx}{\bf{x}}
\nc{\bfy}{\bf{y}}
\nc{\vx}{\vec{x}}
\nc{\vy}{\vec{y}}
\nc{\oB}{\overline{B}}
\nc{\oI}{\overline{I}}
\nc{\oR}{\overline{R}}
\nc{\rar}{\rightarrow}
\nc{\ti}{\times}
\nc{\slsh}{\hskip-5pt/}
\nc{\sm}{Standard~Model~}
\nc{\MP}{M_{\rm Pl}}
\nc{\tp}{t_{\rm Pl}}
\nc{\ave}{\bar{E}}

\renc{\min}{p_{\rm min}}
\renc{\max}{p_{\rm max}}
\nc{\pmin}{p_{\rm min}}
\nc{\pmax}{p_{\rm max}}
\nc{\fo}{f_0}
\nc{\foi}{f_{0,i}\,}
\nc{\fop}{f_0^P}
\nc{\fou}{f_0^U}
\def\sepand{\rule{14cm}{0pt}\and}
\nc{\eff}{{\rm eff}}
\nc{\MT}{M_{\rm T}}
\nc{\ML}{M_{\rm L}}
\nc{\kk}{\vek{k}}
\nc{\pp}{{\rm p}}
\nc{\cb}{critical bubble~}
\nc{\cbs}{critical bubbles~}
\nc{\scb}{subcritical bubble~}
\nc{\scbs}{subcritical bubbles~}
\begin{document}
\topmargin -0.8cm
\headsep 0pt
\topskip -15mm

{\title{{\hfill {{
        }}\vskip 0.5truecm}
{\bf Propagation of Majorana fermions}
\vskip 0.3cm
{\bf in hot plasma}}

 
\author{
{\sc Antonio Riotto$^1$}\\
{\sl NASA/Fermilab Astrophysics Center,}\\
{\sl Fermi National Accelerator laboratory Center,}\\
{\sl Batavia, Illinois 60510-0500, USA}\\
{\sl and}\\
{\sc Iiro Vilja$^{2}$ }\\ 
{\sl Department of Physics, University of Turku }\\
{\sl FIN-20014 Turku, Finland} \\
\sepand
}
\maketitle}
\begin{abstract}
\noindent The properties of Majorana fermions in hot plasma are studied.
One-loop resummed propagator, dispersion relations and their interpretation are 
discussed. It is shown  that particle and hole -like solutions appear as in Dirac/chiral
fermion case. The dispersion relations are, however, crucially different. We  find that, in presence of a large zero temperature bare mass,  hole -like excitations posses a negligible effective mass.  As an example of real  application, we consider the neutralinos in the minimal supersymmetric extension of the  
 standard model and argue that
 for realistic values of the  soft supersymmetry breaking masses the existence of practically massless hole -like excitations have a considerable effect on the thermal properties, {\it e.g.} the thermalization rate,  of
particles interacting with these Majorana excitations.

\end{abstract}
\vfill
\footnoterule
{\small$^1$riotto@fnas01.fnal.gov,  $^2$vilja@newton.utu.fi}
\thispagestyle{empty}
\newpage
\setcounter{page}{1}

The knowledge of the behaviour of a high temperature plasma is crucial to 
explain many puzzles in cosmology, {\it e.g.} the generation of the presently 
observed 
baryon asymmetry in the Universe during the electroweak phase transition 
\cite{ckn}.  It is well-known that the interaction of a  fermion with a plasma 
in thermal equilibrium at  temperature $T$  modifies  the fermionic dispersion 
relation and the poles of the  fermion propagator with respect to the zero 
temperature case \cite{lit1,lit2,lit3}. 
For istance, for an exactly  conserving parity gauge theory at finite 
temperature (like QCD or QED), it has been shown that dispersion relations for
a Dirac fermion are characterized by  two possible solutions of positive 
energy. The addition of a quark to the equilibrium plasma, described by the 
incoherent superposition of many states $|\Phi\rangle$,  produces a fermionic 
excitation (particle) $b^\dagger ({\bf p},\lambda)|\Phi\rangle$ with momentum 
${\bf p}$ and helicity $\lambda$, while  the operator $d (-{\bf p},\lambda)$ 
does not annihilate the ground state as it does at zero temperature. On the 
contrary, the removal of an antiquark $d (-{\bf p},\lambda)|\Phi\rangle$ 
produces a state
with all the same quantum numbers as the particle and   is referred to as a 
hole state (or, more precisely, antiparticle hole state). The energies of 
particles and holes is not the same since there is no combination of parity, 
charge conjugation and time reversal able to relate them. Moreover, in the 
limit of vanishing bare mass $m$ (or $|{\bf p}|\gg m$), particles have  the 
helicity equal to their chirality, while holes have the helicity
opposite to their chirality. Consequently the hole solution propagates with the 
wrong correlation between chirality and helicity. 

Many other different cases have been analyzed  in the literature, but, to  our 
knowledge, attention has been devoted only to the study of the  properties of   
Dirac/chiral fermions propagating in a thermal background. However, 
in  many attractive extensions of the Standard Model (SM) there may appear
Majorana fermions with a non-negligible mass at zero temperature. In some cases 
these Majorana fermions have only chiral interactions. A striking
example characterized by these features is provided by the Minimal 
Supersymmetric extension of the Standard Model (MSSM) \cite{haber} where the 
neutrally charged fermionic superpartners of the boson fields 
present in the SM, called neutralinos,  posses a Majorana nature and may have  
a nonvanishing  bare mass due to soft sypersymmetry breaking
interactions even in the presence of unbroken gauge symmetry. The purpose of 
this Letter is to study the properties of this
kind particles, their resummed propagator, dispersion relation and 
interpretation.
This study is strongly motivated by the recent observation that light stops,  
charginos and neutralinos may play a crucial role in generating the baryon 
asymmetry during the electroweak phase transition \cite{bario}. Since a 
detailed calculation of the  final baryon asymmetry  must 
incorporate the effects of the  incoherent nature of plasma physics on
$CP$-violating observables \cite{hn},  a careful computation of  the  
thermalization rate of supersymmetric particles in the thermal bath by making 
use of improved propagators and including resummation of hard thermal loops is 
called for.  In this paper we will confine ourselves to the inspection of the 
properties of  improved propagators for Majorana fermions having nonvanishing 
bare mass and chiral interactions and 
the full computation of the thermalization rate of supersymmetric particles   will be presented elsewhere \cite{kari}. 

In the MSSM chiral interactions with the surrounding thermal bath produce  
corrections to the inverse propagators of Majorana fermions of the general 
form\footnote{In this paper we will confine ourself to the case of unbroken 
gauge symmetry since the  computation of  the baryon asymmetry is usually made 
by making an expansion of  the propagating Higgs bubble configuration $H(z)$ 
around $H(z)=0$, see M. Carena {\it et al.} in \cite{bario}.}
\be{inverse}
S^{-1}(p) = p_\mu \gamma^\mu + f_\mu \gamma^\mu \gamma^5 - m,
\ee
where $m$ is the zero temperature bare mass. Notice,  the absence of the term 
$f'_\mu\gamma^\mu$ which may be traced back to the nature of Majorana particles. 

The general form given above is easily shown to be valid  by computing, for 
instance, the corrections to the inverse propagators of the $\widetilde{W}^3$- 
and $\widetilde{B}$-neutralinos which are 
mass eigenstates in the unbroken gauge symmetry case\footnote{The case of 
Majorana higgsinos $\widetilde{H}_1^0$ and $\widetilde{H}_2^0$ is more involved 
since they mix even in the unbroken gauge symmetry case due to the presence of 
the term $\mu H_1 H_2$ in the superpotential. This case will be extensively 
considered in \cite{kari}.}. 
Corrections come from the quark-scalar quark-neutralino interactions
\begin{eqnarray}
&-&\sqrt{2}\: g_2\:
\sum_i\: \bar{q}_i \:P_R\:\left[ T_{3i}\:\widetilde{W}^3-{\rm 
tan}\theta_W\:\left(
T_{3i}-e_i\right)\:\widetilde{B}\right]\:\widetilde{q}_{i L}\nonumber\\
&+&\sqrt{2}\: g_2 \:{\rm tan}\theta_W\:\sum_i\:e_i\:\bar{q}_i 
\:P_L\:\widetilde{B}\:\widetilde{q}_{i R} \:+\: {\rm h.c.},
\end{eqnarray}
and the chargino-neutralino-$W^{\pm}$ interaction
\begin{equation}
g\:W^{-}_\mu\:\bar{\widetilde{W}}^3\:\gamma^\mu\:\widetilde{W}.
\end{equation}

At finite temperature the one loop correction function $f_\mu$ is of the form
\be{f}
f_\mu = a(\omega, \vek p) p_\mu + \delta_{0\mu} b(\omega, \vek p)
\ee
with
\be{a}
a(\omega, \vek p) = {m^2(T)\over \vek p^2}\left ( 1 - {\omega\over \vek p}
\ln\left |{\omega + \vek p\over \omega - \vek p}\right | \right )
\ee
and
\be{b}
b(\omega, \vek p) = {m^2(T)\over \vek p}\left [-{\omega\over \vek p} +
\left({\omega^2\over \vek p^2} - 1\right)\frac 12
\ln\left |{\omega + \vek p\over \omega - \vek p}\right | \right ].
\ee 
Here $\omega = p_0,\ \vek p = |${\bf p}$|$ and $m(T)\propto T$ is the
finite temperature plasma mass, $m_{\widetilde{W}^3}^2= (3/16) g^2 T^2$ and 
$m_{\widetilde{B}}^2= (2/9) g_1^2 T^2$ for the example sketched above. 
The  normal fermion mass $m$ is generated
in MSSM by soft supersymmetry breaking terms, usually denoted by $M_2$ and 
$M_1$ for $\widetilde{W}^3$ and $\widetilde{B}$, respectively. 
Note that possible nonchiral interaction corrections would be easy taken in the
account by replacing $p_\mu$ with $p_\mu +  f'_\mu$ where $f'$ would be
of the same form than $f$ but with different  thermal mass $m(T)$.

To obtain the propagator one has simply to invert eq. (\ref{inverse}). The
propagator can in general be given in the form
\be{prop}
S(p) = F + \tilde F \gamma^5 + F_\mu\gamma^\mu + \tilde F_\mu\gamma^\mu\gamma^5
+ F_{\mu\nu} \sigma^{\mu\nu},
\ee
where the sixteen parameters $F,\ \tilde F,\ F_\mu,\ \tilde F_\mu,\ F_{\mu\nu}$
are to be determined. The calculation of propagator is now somehow more
tedious that in the cases of massless Dirac fermions and/or fermions without 
chiral interactions because in these cases some special properties may be used.
In the present case, however, by virtue of tha Lorentz structure of the system
the functions $F_\mu,\ \tilde F_\mu,\ F_{\mu\nu}$ are possible to write in the 
form\footnote{Note that inverting $S^{-1}$ one has no need to take care of the 
fact that at finite temperature the Lorentz symmetry is actually broken to 
simple O(3) symmetry.}
\bea{F's}
F_\mu &=& F_p p_\mu + F_f f_\mu,\\
\tilde F_\mu &=& \tilde F_p p_\mu + \tilde F_f f_\mu,\\
F_{\mu\nu} &=& F_{pf}(p_\mu f_\nu - f_\mu p_\nu) + F_\epsilon
\epsilon_{\mu\nu\alpha\beta}p^\alpha f^\beta,
\eea
where $\epsilon_{\mu\nu\alpha\beta}$ is the usual completely antisymmetric 
tensor and coefficients $F_p,\ F_f,\ \tilde F_p$, $\tilde F_f$, $\tilde F_{pf}$ 
and $F_\epsilon$ are now (pseudo)scalars like $F$ and $\tilde F$. After some
long, but straightforward algebraic manipulations the coefficients read
\bea{coeff}
F &=& - {m(f^2 - m^2 - p^2)\over (p - f)^2(p + f)^2 - m^4},\\
\tilde F &=& 0,\\
F_p &=& {f^2 + m^2 + p^2\over (p - f)^2(p + f)^2 - m^4},\\
F_f &=& - {2(p\cdot f)\over (p - f)^2(p + f)^2 - m^4},\\
\tilde F_p &=& {2(p\cdot f)\over (p - f)^2(p + f)^2 - m^4},\\
\tilde F_f &=& -{ f^2 - m^2 + p^2\over (p - f)^2(p + f)^2 - m^4},\\
F_{pf} &=& 0,\\
F_\epsilon &=& {m\over (p - f)^2(p + f)^2 - m^4}.
\eea
One can establish that in the special cases $m = 0$ and/or $f_\mu = 0$ the 
propagator coincides with the known ones.

From the denominator of the propagator one can read out the dispersion
relation (naturally coinciding with $\det S^{-1} = 0$)
\be{dispersion}
(p - f)^2(p + f)^2 - m^4 = 0.
\ee
This equation appears to have two different positive energy solutions
$\omega_\pm(\vek p)$. Even in 
massless limit $ m \goto 0$ only two different positive energy solutions exist,
because it is easy to show that  $(p - f)^2 > 0$ for $\omega > 0$. Note that 
only 
the positive energy solutions are physically independent, because for Majorana
particle the antiparticle, corresponding the negative energy solution coincides
with the particle itself. 

The zero momentum limit $ \vek p \goto 0$, however, can be solved exactly. It 
appears that the effective masses are given by ($m_\pm > 0$)
\be{effmass}
m_\pm^2 \equiv \omega^2_\pm(0) = \frac 12 \left (\sqrt{ m^4 + 4\, m^4(T)} 
\pm m^2\right ).
\ee
Here we make the crucial observation that in the limit of large bare mass, 
$m\gg m(T)$, the hole excitation effective mass $m_{-}$ becomes much smaller 
than $m(T)$ and the temperature, $m_{-}\ll m(T)< T$. 
Expression  (20) could be compared to the corresponding one for the  effective 
masses of  pure Dirac case (with nonvanishing bare mass $m$) where
$ f_\mu = 0$, but with a $\tilde f_\mu \gamma^\mu$ -term:  effective masses 
are given by $(\omega^{{\rm Dirac}}_\pm (0))^2 = \frac 12 [m^2 + 2 m^2(T) \pm
\sqrt{m^4 + 4 m^2 m^2(T)}]$. The reader should remember that SM Dirac  fermions
do not posses a bare mass in the case in which the $SU(2)_L\otimes U(1)_Y$ 
gauge symmetry is unbroken and their effective mass at finite temperature 
reduces to the the plasma mass $m(T)$. On the contrary, supersymmetric Majorana 
fermions receive a bare mass from supersymmetry breaking even in the case of 
unbroken gauge symmetry and are characterized by the novel feature
that the hole -like excitation may posses a very small effective mass. This 
property will necessarily affect the kinematics involved in the computation of 
the thermalization
rate of the degrees of freedom interacting with these hole -like excitations.

It can be also proved that for $m\ne 0$ the derivative of $\omega$ with respect
to spatial momentum is zero, {\it i.e.} ${\partial \omega_\pm(\vek p)\over 
\partial \vek p} = 0$.
This is in contrast with the knowledge about the massless Dirac and chiral
fermions, where it is possible to show that ${\partial \omega(\vek p)\over 
\partial\vek p} = \pm \frac 13$. It can be, however, shown that this peculiar 
property
is not general but is closely related to vanishing of mass $m = 0$. Thus
Majorana case represents  no exemption in this sense. Indeed, at the massless 
limit $m = 0$ the solutions of Eq. (\ref{dispersion}) coincides to the known 
ones with ${\partial \omega_\pm(\vek p)\over \partial \vek p} = \pm \frac 13$.

Although no simple formula for the solutions can be given, asymptotic formulas 
for small and large $\vek p$ are possible to calculate. For small momenta
$\vek p \sim 0$ they read
\be{smallp}
\omega_\pm(\vek p) \simeq m_\pm + K_\pm \vek p^2,
\ee
where
\be{K}
K_\pm \equiv \left.\frac 12 {\partial^2\omega_\pm\over \partial \vek 
p^2}\right|_{\vek p 
= 0} = {1 + \frac 49 \left ({m(T)\over m_\pm}\right )^4 - \frac 59
\left ({m(T)\over m_\pm}\right )^8\over 2 m_\pm \left [ 1 -
\left ({m(T)\over m_\pm}\right )^8\right ]}.
\ee
From this we can read out that there exists a critical value for the ratio
$m/m(T)$, 
\be{crit}
\left.{m\over m(T)}\right|_{{\rm crit}} = \left ({4\over 3\sqrt 5}\right )^{1/2}
\simeq 0.772
\ee
such that for values smaller that it, $\omega_{-}(\vek p)$ has a minimum for
some value of $\vek p$. This phonomenon is similar than occurs in Dirac field 
case and can be undestood because at the limit $ m \goto 0$ the solution
of Majorana case tends smoothly towards the Dirac one. However the derivative 
of $\omega_\pm$ at $\vek p = 0$ is not continuous at the same limit. This 
reflects the fact that free field theory contains naturally the mass term, too.
So from the point of view of naturalness $m = 0$ is a kind of pathological 
case. In Figs. 1 and 2 the dispersion relations are schematically given in two
cases $ m/m(T) = 1 > \left.{m\over m(T)}\right|_{{\rm crit}}$ (Fig. 1) and 
$ m/m(T) =1/4 < \left. {m\over m(T)}\right|_{{\rm crit}}$ (Fig. 2).

For large $\vek p$ values the functions $\omega_\pm$ are asymptotically given by
\be{largepm}
\omega_{-}(\vek p) \sim \vek p \left( 1 + 2\: {\rm e}^{-1 - {\vek p^2\over 
m^2(T)} \sqrt{\left ({m\over m(T)}\right )^4 + 4}}\right)
\ee
and
\be{largepp}
\omega_+(\vek p) \sim \vek p + {\sqrt{m^4 + 4\, m^4(T)}\over 2 \vek p} - 
{M^4\over 2 \vek p^3}\ln {4 \:\vek p^2\over \sqrt{m^4 + 4\, m^4(T)}}.
\ee
These relations coincide with the ones known in the literature for the Dirac 
fermion case \cite{lit1}
at the limit $m \goto 0$. It should be noted that, for non-zero $m$, 
$\omega_{-}$ tends towards the asymptote $\omega = \vek p$ even faster that
the massless case, and therefore these modes also disappears from the spectrum
very fast when momentum increases.

Let us now  discuss the nature of the thermal excitations for the Majorana 
fermions. To do so, let us first remind the reader what happens for a Dirac 
fermion.  At zero temperature, the ground state is the vacuum. Only four 
different excitations may be created from the vacuum: $b^\dagger ({\bf 
p},\lambda)|{\rm vac}\rangle$ (fermion) and $d^\dagger ({\bf p},\lambda)|{\rm 
vac}\rangle$ (antifermion) each with two values of helicity $\lambda$. At finite
temperature, the ground state of the plasma contains a large number of 
fermion-antifermion pairs, which are not the virtual present at $T=0$ because 
of quantum fluctuations, but real \cite{lit1}. In such a case there are eight 
different operators that are able to create elementary excitations: 
$B_p^\dagger({\bf p},\lambda)$ and $B_h^\dagger({\bf p},\lambda)$, which are a 
linear combinations of the usual $T=0$ fermion creation operator $b^\dagger(
{\bf p},\lambda)$ and antifermion annihilation operator $d(-{\bf p},\lambda)$,
and $D_p^\dagger(-{\bf p},\lambda)$ and 
$D_h^\dagger(-{\bf p},\lambda)$, which are a linear combinations of the usual 
$T=0$ antifermion creation operator $d^\dagger(-{\bf p},\lambda)$ and fermion 
annihilation operator $b({\bf p},\lambda)$. The features of the hole states 
become clear in the massless limit when, for instance, $B_h^\dagger({\bf 
p},\lambda)\equiv
d(-{\bf p},\lambda)$. Consider an antifermion in the plasma with momentum $
-{\bf p}$ and spin polarized along the direction $-\lambda \hat{{\bf p}}$. It 
has chirality $\lambda$. When applied to the $T\neq 0$ ground state $d(-{\bf 
p},\lambda)$ produces the removal of this antifermion and the creation of a 
hole with momentum ${\bf p}$ and spin along  $\lambda\hat{{\bf p}}$. The hole 
has, therefore, chirality $-\lambda$, but helicity $\lambda$. The hole has the 
wrong correlation between chirality and helicity \cite{lit1} and it is referred 
to as an antifermion hole. 

In the Majorana case, since at $T=0$ particles and antiparticles coincide,  the 
operators $b^\dagger({\bf p},\lambda)$
and $d^\dagger(-{\bf p},\lambda)$ have to be identified and there is no longer 
any distinction between a fermion and an antifermion. In other words, when 
$B_p^\dagger({\bf p},\lambda)\equiv D_p^\dagger(-{\bf p},\lambda)$ is applied 
to the ground state, it generates a Majorana fermion with spin along 
$\lambda\hat{{\bf p}}$ and helicity $\lambda$, while $B_h^\dagger({\bf 
p},\lambda)\equiv D_h^\dagger(-{\bf p},\lambda)$, when   applied to the ground 
state, it generates a Majorana hole (or Majorana fermion hole)  with spin along 
$-\lambda\hat{{\bf p}}$ and the wrong helicity  $\lambda$. Only 
the positive energy solutions associated to the operators $B_{p,h}^\dagger({\bf 
p},\lambda)$ are physically independent, because for Majorana
particle the antiparticle, corresponding the negative energy solution coincides
with the particle itself. 

In conclusion, in this letter we have studied the structure of Majorana 
fermions with chiral
interactions in hot plasma. An example of such theory is provided by the  
MSSM where neutralinos are such fields. For them, the value of the 
ratio $m/m(T)$ is usually larger than the critical value  $\left.{m\over 
m(T)}\right|_{{\rm crit}} \simeq 0.772$, 
even though smaller values are not excluded.  We have also found that 
for large $m$ the Majorana hole excitation effective mass 
$m_{-}$ may be very small,  $m_{-} \ll T$, and 
therefore it may have remarkable effect on, {\it e.g.}, the  thermalization rate
(and thus on the coherence length) of particles interacting with them. In MSSM
the usually adopted values of the soft supersymmetry breaking masses  are 
larger than the corresponding thermal masses and therefore the smallness of 
$m_{-}$ may drastically affect the properties of the other 
degrees of freedom present in the hot plasma \cite{kari}.

\vspace{1cm}
{\bf Acknowledgement}

We would like to thank all members of the High Energy Physics group at ICTP for 
the kind hospitality and friendly environment when this work was initiated. In 
particular, we are grateful to Alexei Yu. Smirnov for enlightening discussions.
AR would also like to thank Eleonora Riotto for entertaining conversations. AR 
is supported by the DOE and NASA under Grant NAG5--2788.

\newpage

\newpage
%
\noindent {\Large{\bf Figure captions}}
\vskip .5truecm
\noindent Figure 1. Dispersion relations of Majorana particles with 
$m/m(T) = 1$.
\vskip .5truecm
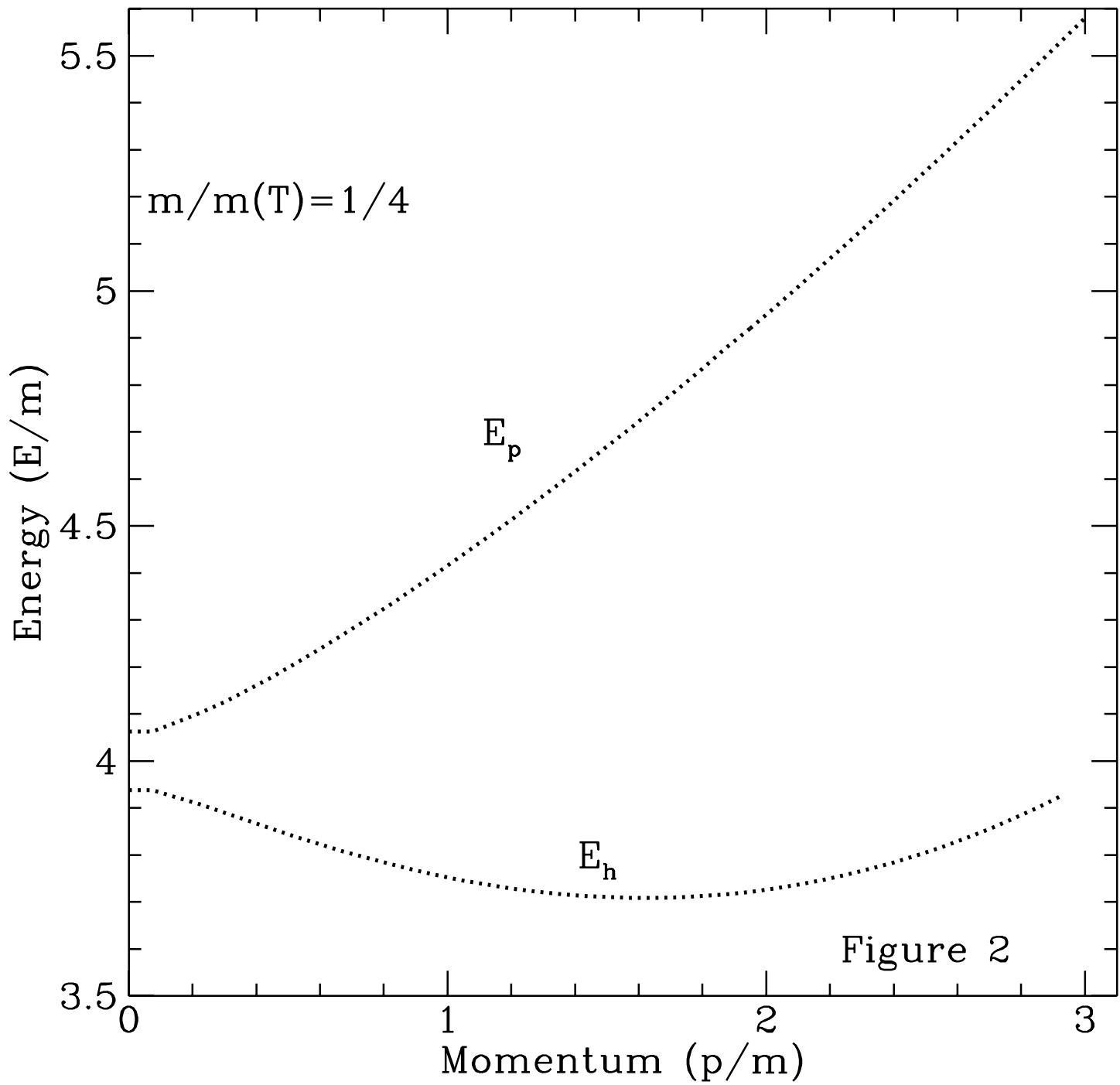
\noindent Figure 2. Dispersion relations of Majorana particles with 
$m/m(T) = 1/4$.

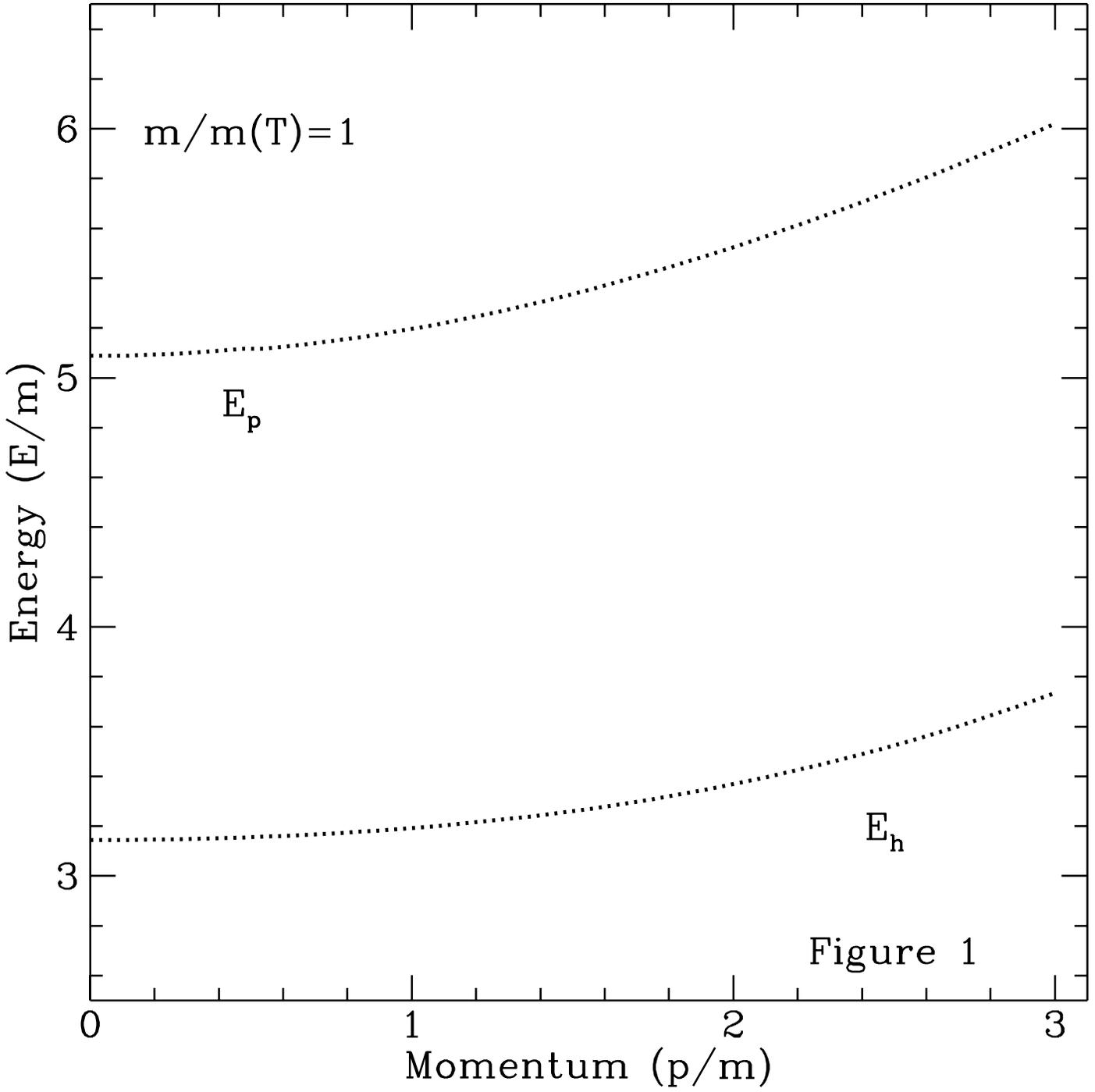
\end{document}